\documentclass[12pt]{article}
\usepackage[numbers]{natbib}
\usepackage{a4}
\usepackage{amsmath}
\usepackage{amssymb}
\usepackage{latexsym}
\usepackage{epsfig}
\usepackage{color}
\usepackage{ulem}
\usepackage{psfrag}
\def \d{{\mathrm{d}}}

\def \D{{\mathrm{D}}}

\def \pd{\partial}

\def \iner{\rfloor}
\def \hodge{{}^\star}

\def \e{{\mathrm{e}}}

\def \tl#1{\overset{\kern 1pt\circ}{#1}}
\def \TL#1{\overset{\kern -3pt \circ}{#1}}
\def \TLL#1{\overset{\kern -7pt \circ}{#1}}

\def \AA{{\cal{A}}}

\definecolor{dark-green}{rgb}{0,0.7,0}
\definecolor{dark-blue}{rgb}{0,0.2,0.5}
\definecolor{med-blue}{rgb}{0,0.7,1}
\definecolor{mblue}{rgb}{0,0.2,1}
\definecolor{cnc}{rgb}{0.8,0,0}
\definecolor{light-red}{rgb}{1,0.8,0.8}
\definecolor{dark-yellow}{rgb}{1,0.8,0}
\definecolor{light-blue}{rgb}{0.8,0.9,1}
\definecolor{verylight-blue}{rgb}{0.93,0.95,1}
\definecolor{light-yellow}{rgb}{1,0.9,0.8}
\definecolor{grey}{gray}{0.88}

\definecolor{new-green}{rgb}{0.5,0.5,0.6}

\def\a{\alpha}
\def\b{\beta}
\def\g{\gamma}
\def\de{\delta}
\def\e{\epsilon}

\def\vt{\vartheta}

\begin{document}

\title{{\bf Cartan's spiral staircase in physics and, in particular,
    in the gauge theory of dislocations\footnote{Dedicated to
      Professor Peter Mittelstaedt on the occasion of his 80th
      birthday.}}}

\author{Markus Lazar$^\text{a,b,}$\footnote{E-mail:
    lazar@fkp.tu-darmstadt.de} \;and Friedrich W.\
  Hehl$^\text{c,d,}$\footnote{E-mail:
    hehl@thp.uni-koeln.de}
\\ \\
${}^\text{a}$
        Emmy Noether Research Group,\\
        Department of Physics,\\
        Darmstadt University of Technology,\\
        Hochschulstr. 6,\\
        64289 Darmstadt, Germany\\
${}^\text{b}$ 
Department of Physics,\\
Michigan Technological University,\\
Houghton, MI 49931, USA\\
${}^\text{c}$
Institute for
 Theoretical Physics, University of Cologne,\\ 50923 K\"oln, Germany\\
${}^\text{d}$
Dept.\ of Physics \& Astronomy, University of Missouri,\\ Columbia, MO
 65211, USA}

\date{23 February 2010, {\it file Cartan28.tex}}
\maketitle
\begin{abstract}
  In 1922, Cartan introduced in differential geometry, besides the
  Riemannian curvature, the new concept of {\it torsion}. He
  visualized a homogeneous and isotropic distribution of torsion in
  {\it three dimensions} (3d) by the ``helical staircase'', which he
  constructed by starting from a 3d Euclidean space and by defining a
  new connection via helical motions. We describe this geometric
  procedure in detail and define the corresponding connection and the
  torsion. The interdisciplinary nature of this subject is already
  evident from Cartan's discussion, since he argued---but never
  proved---that the helical staircase should correspond to a continuum
  with constant {\it pressure} and constant internal {\it torque}. We
  discuss where in physics the helical staircase is realized: (i)
  In the continuum mechanics of Cosserat media, (ii) in (fairly
  speculative) 3d theories of gravity, namely a) in 3d Einstein-Cartan
  gravity---this is Cartan's case of constant pressure and constant
  intrinsic torque--- and b) in 3d Poincar\'e gauge theory with the
  Mielke-Baekler Lagrangian, and, eventually, (iii) in the gauge field
  theory of {\it dislocations} of Lazar {\it et al.}, as we prove for
  the first time by arranging a suitable distribution of screw
  dislocations. Our main emphasis is on the discussion of dislocation
  field theory.
\end{abstract}

\bigskip

Keywords: Cartan's torsion, differential geometry, dislocations,
Cosserat continuum, Einstein-Cartan theory, 3-dimensional theories of
gravitation

\section{Introduction: Homogeneous and isotropic torsion in three dimensions}

\subsection{Cartan's original idea}

In 1922, when \'Elie Cartan \cite{Cartan1922,Cartan} analyzed
Einstein's general relativity theory (GR), he introduced in this
context the concept of torsion into differential geometry.  Thereby he
generalized Riemannian geometry, dominated by the metric tensor
$g_{ij}=g_{ji}$ and the Riemannian curvature tensor
$\widetilde{R}_{ijk}{}^\ell$, to Riemann-Cartan (RC) geometry carrying
a generalized curvature ${R}_{ijk}{}^\ell$ and an additional
fundamental third rank tensor $T_{ij}{}^k=-T_{ji}{}^k$, which was
named torsion by Cartan; here $i,j,\dots$ are coordinate indices
running either over $1,2,3$ (space) or over $0,1,2,3$ (spacetime).

Whereas it is simple to visualize say a 2-dimensional (2d) Riemannian
space as a curved 2d surface imbedded in a (flat) 3d Euclidean space,
no simple picture lends itself to a visualization of a space with
torsion. Still, already in his first publication on the subject,
Cartan \cite{Cartan1922}, starting from 3d Euclidean space, gave a
prescription of how to arrive at a specific 3d space with homogeneous
and isotropic torsion. We refer to this space as {\it Cartan's spiral
  staircase} for reasons that will become clear in the next two
paragraphs. This construction is largely forgotten\footnote{See,
  however, Garc{\'\i}a {\it et al.}~\cite{Garcia} and Hehl and Obukhov
  \cite{HO07}.}  in spite of being quite helpful in explaining the
characteristic features of a simple space with torsion.

The idea of Cartan was the following: Take a point $A$ of a 3d
Euclidean space $E$ in Cartesian coordinates, as it is depicted in
Figure 1. Consider a neighboring point $A'$. The vector linking $A$
with $A'$ will be denoted by $\stackrel{\longrightarrow}{A{A'}}$. {\it
  Rotate} now the triad in $A'$ in accordance with the vector
$\vec{\omega}:=\lambda\stackrel{\longrightarrow}{A{A'}}$ in the right
hand-sense, where $\lambda$ is a prescribed constant. The new rotated
triad serves as a basis for the space $F$ with torsion: a vector in
$F$ at $A'$ is said to be parallel to a vector at $A$, if its
components in $A$ with respect to the local triad are the same as in
$A'$ with respect to the {\it rotated} triad. Whereas the Euclidean
connection $\widetilde{\Gamma}$ is zero with respect to the triads in
$E$, the new Riemann-Cartan connection $^\xi\Gamma$ vanishes with
respect to the rotated triads. This new ``helical'' connection
$^\xi\Gamma$ carries a non-Riemannian piece that is proportional to
the torsion.  The space $F$ is like our ordinary space as viewed by an
observer whose perception has been twisted \cite{Cartan1922}.

Now, the vector $\vec{\omega}$ can be decomposed into its
components $\omega^1,\omega^2,\omega^3$, that is, into rotations around the
$x$-, $y$-, and $z$-axis, see Figure 1. Accordingly, if $A'$ first
coincides with $A$ and is then shifted further and further away from
$A$, then the triad along each of the three axes undergoes a helical
motion, that is, it is like going up a spiral staircase along each of
the axes.

In Section 2, following the prescriptions of Cartan, we will
construct, with the help of the calculus of exterior differential
forms, the Riemann-Cartan connection $^\xi\Gamma$ of the spiral
staircase $F$. In this way, we can put Cartan's idea in a very
succinct form. We will determine the two-forms of torsion $T^\a$ and
of curvature $R^{\a\b}$ of the spiral staircase. This will allow us to
understand the 3d Riemann-Cartan space under consideration.

\begin{figure}[h!]\label{Fig1}
\begin{center}
    \psfrag{A}{$x$}
    \psfrag{B}[c][c][1]{$y$}
    \psfrag{C}[c][c][1]{$z$}
    \psfrag{D}[c][c][1]{$A$}
    \psfrag{E}[c][c][1]{\textcolor{red}{$\omega^1$}}
    \psfrag{F}[c][c][1]{\textcolor{red}{$\omega^2$}}
    \psfrag{G}[c][c][1]{\textcolor{red}{$\omega^3$}}
    \psfrag{H}[c][c][1]{\textcolor{red}{$\omega^k$}}
   \psfrag{I}[c][c][1]{$A'$}
   \psfrag{J}[c][c][1]{The new \textcolor{red}{red} connection determines~$F$}
    \psfrag{K}[c][c][1]{\underline{\'E.~Cartan's Construction (1922)}}
    \psfrag{L}[c][c][1]{Euclidean space~$E$}
    \psfrag{M}[c][c][1]{\textcolor{red}{RC space~$F$}}
   \includegraphics{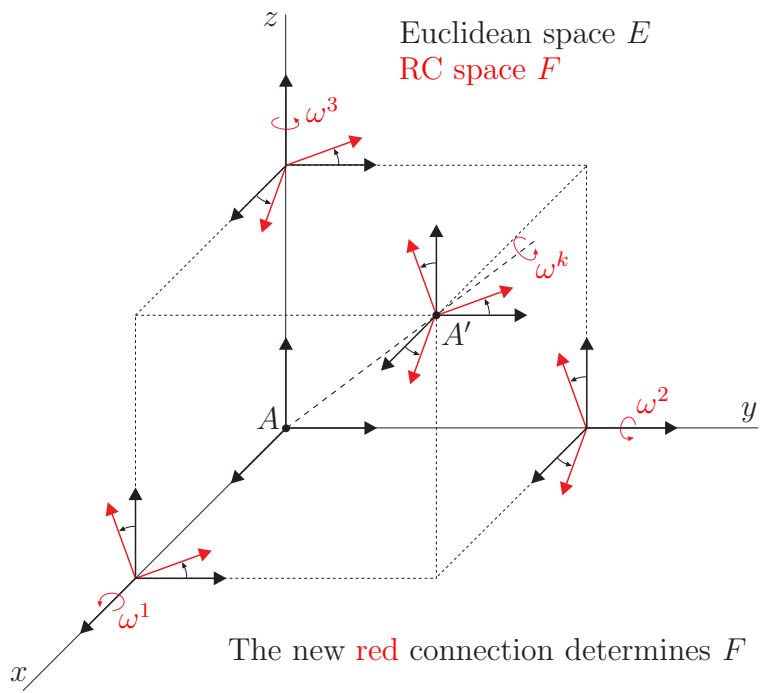}
\caption{{\it Cartan's construction (1922) of a 3-dimensional space
    with homogeneous and isotropic torsion: the spiral staircase}}
\end{center}
\end{figure}

\subsection{Cosserat medium}

In his original article, Cartan \cite{Cartan1922} argued---and now we
are back to physics---that the space $F$ of the spiral staircase
corresponds to a medium with constant {\it pressure} $p$ and constant
{\it internal torque} $\tau$. As is clear from his acknowledgment,
Cartan was influenced in his investigations by the work of the
brothers Cosserat \cite{Cosserat1909} on continuum mechanics. In
classical continuum mechanics \`a la Euler and Cauchy---to name two
main figures of the orthodoxy---a medium with internal torque does not
exist, since the classical medium can only support (force-)stresses
$\sigma_i{}^k$ (specifically with pressure $p:=\frac 13\sigma_k{}^k$),
but no internal torques.

However, already Voigt \cite{Voigt1887} had introduced the new concept
of a (spin) moment stress tensor $\tau_{ij}{}^k=-\tau_{ji}{}^k$, which
can exist in a suitable medium in addition to the (force) stress
tensor $\sigma_i{}^k$. Then, specifically an internal torque
$\tau:=\frac 12\epsilon^{ijk}\tau_{ijk}$ can be defined, with
$\epsilon^{ijk}=\pm1,0$ as totally antisymmetric Levi-Civita
symbol. The notion of a (spin) moment stress tensor has been brought
to fruition in the work of the brothers Cosserat.\footnote{For insight
  into more recently developed continuum theories with microstructure
  and with generalized stresses, see, for example,
  \cite{Jaun,Maugin:1993,Hyperstress,Eringen99,Maugin:2003a,Maugin:2003b,LM07};
  also the review of Neff \cite{Neff} is very readable.}

To get an intuitive feeling for their type of approach, we can start,
as Cartan did, with Euclidean geometry and its fundamental notions of
{\it translation} and {\it rotation}. On the one hand, these two
geometrical notions are---via a suitable variational principle---
intrinsically related to the static notions of force stress and spin
moment stress, on the other hand, if one restricts the validity of
Euclidean geometry only to local neighborhoods (``gauging of the
Euclidean group''), one ends up, guided by Cartan, with a
Riemann-Cartan geometry with torsion and curvature.

This brings us back to Cartan's medium with constant pressure $p$ and
constant internal torque $\tau$ as image of his spiral staircase:
Apparently a Cosserat medium, generalizing the medium of classical
elasticity, fluid mechanics, etc., is appropriate for this
purpose. Accordingly, Cartan's introduction of the torsion has as a
consequence a more general conception of a continuum than the one
taken for granted in classical continuum mechanics.

As far as we aware, nobody considered so far the implications of the
spiral staircase to a Cosserat continuum. We will describe our
corresponding results that the spiral staircase produces 
constant hydrostatic pressure and constant internal torque 
in a linear, incompatible isotropic Cosserat medium
in Section 3.

\subsection{Three-dimensional gravity}

Let us recall that Cartan was in the process of analyzing GR, that is,
a gravitational theory. In the course of these investigations, he
developed the skeleton of a new, slightly generalized theory of
gravity. This four-dimensional theory of gravitation, to which Cartan
laid the foundations, was worked out by Sciama and Kibble around 1961,
see \cite{PRs,Trautman}, and is called Einstein-Cartan (EC) theory of
gravity. It is one of the very few viable theories of gravity.

In the meantime, however, also a somewhat speculative
three-dimensional (3d) EC-type model of gravity, with a Lagrangian
proportional to the curvature scalar of the RC-space, has been
proposed. This 3d EC-model has an {\it exact solution} with a geometry
described by the spiral staircase and {\it matter} of the Cosserat
type carrying constant pressure and constant spin moment stress
(torque). We believe that it is this solution that Cartan described in
words. The corresponding derivations and the details will be presented
in Section 4.

Moreover, in the 3d framework there also exist gravitational models
with Lagrangians {\it quadratic} in torsion and curvature (Poincar\'e
gauge models). The most general one of these is a model of Mielke \&
Baekler \cite{MB91,BMH92}, see also Mielke {\it et al.}\ \cite{Mielke:2003}
and Klemm {\it et al.}\ \cite{Klemm} in which Cartan's helical staircase
occurs and plays an important role. The Mielke-Baekler model has a
so-called BTZ (Ba{\~n}ados, Teitelboim, Zanelli) black hole {\it with}
torsion as an exact vacuum solution \cite{Garcia}. Also in Section 4
we show that a subcase of the vacuum BTZ-solution with torsion (namely
for $\Lambda_{\rm eff}=0$) carries the torsion of the spiral
staircase---and for vanishing mass and angular momentum $M=0,\,J=0$,
also its connection $^\xi\Gamma^{\a\b}$. In contrast to the
EC-solution mentioned above, it is an exact {\it vacuum} solution and,
accordingly, was outside the scope of Cartan in 1922, as we shall see
in Section 4.

\subsection{Gauge theory of dislocations}

Let us now turn to an important point of our investigations: During
the 1950s it became clear that {\it crystal dislocations} can be
described by Cartan's torsion~\cite{Kondo,Kroener60}, 
basically since dislocations, similar
as torsion, can break open infinitesimally small parallelograms, in
this way inducing a {\it closure failure} of the parallelogram; for
reviews, see Kr\"oner \cite{Kroener1980} and, furthermore, Ruggiero
and Tartaglia \cite{RT}. Since dislocation lines are discrete objects,
it is helpful to consider in this context a ``continuized crystal''
\cite{Kroener1986}.

During the last years a successful gauge field theory of dislocations
has been developed by Lazar~\cite{Lazar00,Lazar02,Lazar09b} and Lazar and
Anastassiadis \cite{LA08,LA09,LA09b}; for different versions, see
\cite{Katanaev:1992kh,Katanaev:2004xq,Malyshev,Kleinert} and the
reviews \cite{Puntigam:1996,Kleman,Kleman:2009}. Within this theory,
the equivalence between torsion and dislocation density plays a
leading role. Then immediately the question comes up whether Cartan's
spiral staircase is an {\it exact solution} of Lazar's gauge theory of
dislocations and whether one can find, indeed, constant pressure and
constant internal torque in the corresponding medium.

It appears intuitively clear that the spiral staircases along the
three Cartesian axes---with the same pitch!---must be implemented by
three forests of parallel {\it screw dislocations}; the forests should
be perpendicular to each other and each of them be of the same
dislocation density, that is, their respective Burgers vectors should
be the same and constant, see \cite{HK}. This distribution of screw
dislocations should provide a constant torsion of the twisted type
that, by means of the constitutive laws, should induce constant
pressure and constant internal torque---provided Cartan's intuition
was right and Lazar's theory appropriate.

We prove in Section 5 that with Lazar's highly non-trivial gauge
theory of dislocations we can derive, in linear approximation, the
constant pressure and the constant internal torque for the first time
in a realistic setting.

In Section 6 we discuss our results and compare them with the literature.

\section{The differential geometry of Cartan's spiral staircase}

\subsection{Mathematical framework, conventions\protect\footnote{A systematic
    presentation of our formalism can be found, for example, in
    Ref. \cite{Birkbook}. }}

The geometrical arena for our considerations  consists of a
three-dimensional manifold $M$ together with a local Euclidean {\it
  metric} $g$. At each point we introduce a {\it coframe} field
$\vt^\a $, with the anholonomic (or frame) indices
$\a,\b,...=\hat{1},\hat{2},\hat{3}$, and, dual to it, the {\it frame}
field $e_\b$, with $e_\b\rfloor\vt^\a=\delta_\b^\a$, where $\rfloor$
denotes the interior product. In a trivial orthonormal coframe $
\vt^\a=\delta_i^\a \,\d x^i$, we have
\begin{align}
g=g_{\a\b}\, \vt^\a \otimes\vt^\b,\qquad
g_{\a\b}\stackrel{*}{=}\mbox{diag}(1,1,1)\,;
\end{align}
the corresponding trivial frame field is $
e_\b=\delta_\b^j\partial_j$. Thus, more explicitly, we have
\begin{align}
  \vt^{\hat{1}}= \d x^1\,,\; \vt^{\hat{2}}= \d x^2\,,\;
  \vt^{\hat{3}}= \d x^1\,,\quad
  e_{\hat{1}}=\partial_1\,,\;e_{\hat{2}}=\partial_2\,,\;
  e_{\hat{3}}=\partial_3\,,
\end{align}
the holonomic (or coordinate) indices $i,k,...$ run over $1,2,3$.
A $\vartheta$-basis for 0-, 1-, 2-, and 3-forms is
$\left\{1,\vt^\a,\,\vt^{\a\b}:=\vt^\a\wedge\vt^\b,\,\vt^{\a\b\g}:=
\vt^\a\wedge\vt^\b\wedge\vt^\g\right\}$, the Hodge dual $\eta$-basis for
3-,2-,1-, and 0-forms is specified by
\begin{align}
  \eta&:=\hodge 1=\frac{1}{3!}\, \eta_{\a\b\g}\, \vt^{\a\b\g}\,,\\
  \eta_\a&:=\hodge\vt_\a=e_\a\rfloor\eta=\frac{1}{2}\,
  \eta_{\a\b\g}\, \vt^{\b\g}\,,\\
  \eta_{\a\b}&:=\hodge(\vt_{\a\b})=e_\b\rfloor\eta_\a= \eta_{\a\b\g}\, \vt^\g\,,\\
  \eta_{\a\b\g}&:=\hodge(\vt_{\a\b\g})=e_\g\rfloor\eta_{\a\b}  \,,
\end{align}
where $\hodge$ denotes the Hodge star and
$\eta_{\a\b\g}:=\sqrt{{\text{det}}\, (g_{\mu\nu})}\,
\epsilon_{\a\b\g}$, with $\epsilon_{\a\b\g}$ as the totally
antisymmetric Levi-Civita symbol with $\pm1,0$.

This formalism can be straightforwardly generalized to $n$ dimensions
and to a Lorentzian metric with, in $n=4$,
$g_{\a\b}\stackrel{*}{=}\mbox{diag}(-1,1,1,1)$.

In the case of the spiral staircase, see Figure 1, we have an
underlying three-dimensional Euclidean space $E$ with {\it flat}
metric (that is, vanishing Riemann curvature) and, accordingly, with a
Euclidean connection 1-form
$\Gamma^{\a\b}=-\Gamma^{\b\a}=\Gamma_i{}^{\a\b} \d x^i$ that has
vanishing torsion and vanishing curvature.

\subsection{Cartan's prescription}

Cartan introduced, besides this Euclidean connection $\Gamma^{\a\b}$,
for the space $F$ a non-Euclidean {helical} {\it Riemann-Cartan
  connection} 1-form,
\begin{align}
\label{conn}
  ^\xi\Gamma^{\a\b}=-^\xi\Gamma^{\b\a} ={}^\xi\Gamma_i{}^{\a\b}\d x^i
  ={}^\xi\Gamma_1{}^{\a\b}\d x^1+{}^\xi\Gamma_2{}^{\a\b}\d x^2+
  {}^\xi\Gamma_3{}^{\a\b}\d x^3\,,
\end{align}
in the following way, see Figure 1: If we consider the non-Euclidean
parallel displacement along the $x^1$-axis, then, according to
Cartan's recipe, the corresponding orthonormal coframe rotates in the
positive sense around the $x^1$-axis with the angle
$\omega=\omega^{23}=-\omega^{32}$ per unit length. The analogous
prescription applies to the parallel displacements along the
$x^2$-axis and the $x^3$-axis. Then only the following connection
components, up to their antisymmetry, are non-vanishing and equal:
\begin{align}\label{conncomp1}
  ^\xi\Gamma_1{}^{\hat{2}\hat{3}}= {}^\xi\Gamma_2{}^{\hat{3}\hat{1}}=
  {}^\xi\Gamma_3{}^{\hat{1}\hat{2}}=\omega\,.
\end{align}
We transform the coordinate indices into frame indices and find,
\begin{align}
\label{conncomp2}
{}^\xi\Gamma^{\g\a\b}:=g^{\g\delta}e^i{}_\delta\, {}^\xi\Gamma_i{}^{\a\b}
= \omega\eta^{\g\a\b}\,,
\end{align}
with $e_\a=e^i{}_\a\partial_i\stackrel{*}{=}\delta^i_\a\partial_i$ and
$g_{\a\b}=\mbox{diag}(1,1,1)$.
In an anholonomic basis, the connection 1-form reads:
\begin{align}
  \label{Kon-anh} {}^\xi\Gamma^{\a\b}={}^\xi\Gamma^{\g\a\b}
  \vt_\g=\omega\eta^{\a\b\g}\vt_\g=\omega\,\eta^{\a\b}\,.
\end{align}

For the coframe we have, see Figure 1, $\vt^{\hat{1}}=\d x^1\,,\vt^{\hat{2}}=\d x^2\,,\vt^{\hat{3}}=\d x^3$, which is a trivial coframe
\begin{align}\label{coframe}
\vt^\a=\delta^\a_i \, \d x^i\, .
\end{align}
Accordingly, the torsion 2-form is constant and only its axial piece
survives:
\begin{align}
\label{torsion}
  T^\a:=\D\vt^\a=\d\vt^\a+ {}^\xi\Gamma_\b{}^\a\wedge
  \vt^\b=\omega\eta^{\a\b\g}\vt_\b\wedge\vt_\g=2\omega\,\eta^\a\,.
\end{align}
For the Riemann-Cartan curvature 2-form we find\footnote{In
  \cite{HO07} the curvature carries a different sign, since there the
  Lorentzian signature $-++$ was used.}
\begin{align}
\label{curv}\nonumber
  R_\a{}^\b&:=\d\, {}^\xi\Gamma_\a{}^\b- {}^\xi\Gamma_\a{}^\g\wedge
  {}^\xi\Gamma_\g{}^\b
  =-\omega^2
  \eta_\a{}^{\delta\g}\eta^{\varepsilon\b}{}_\g\,\vt_\delta\wedge
  \vt_\varepsilon
\nonumber\\
  &= -\omega^2\left(\delta_\a^\varepsilon g^{\delta\b}-\delta_\a^\b
    g^{\delta\varepsilon} \right)\vt_\delta\wedge \vt_\varepsilon=
  \omega^2\,\vt_\a\wedge\vt^\b\,\qquad\mbox{or}\nonumber\\
R^{\a\b}&=\omega^2\, \vt^{\a\b}\,.
\end{align}

Alternative to the torsion 2-form $T^\a$, we can define the
contortion 1-form $K_{\a\b}=-K_{\b\a}$ either implicitly by
\begin{equation}\label{contortion}
T^\a=:K^\a{}_\b\wedge\vt^\b\,
\end{equation}
or explicitly by
\begin{equation}
\label{contortion1}
K_{\alpha\beta}= e_{[\alpha}\rfloor T_{\beta ]}
-\frac{1}{2}\, ( e_{\alpha}\rfloor e_{\beta}\rfloor
T_{\gamma})\vartheta^{\gamma} = 2e_{[\alpha}\rfloor T_{\beta ]}
-\frac{1}{2}\, e_{\alpha}\rfloor e_{\beta}\rfloor
(T_{\gamma}\wedge\vartheta^{\gamma})\,.
\end{equation}
Simple algebra yields for Cartan's spiral staircase
\begin{equation}
\label{CON}
K_{\a\b} =-\omega\, \eta_{\a\b}\,.
\end{equation}

In order to isolate the Riemannian part,
we decompose the Riemann-Cartan connection
into the Levi-Civita (or Christoffel)
connection $\widetilde{\Gamma}^{\a\b}$ and the
contortion $K^{\a\b}$ as follows
\begin{align}
\label{Con-deco}
{}^\xi\Gamma^{\a\b}=\widetilde{\Gamma}^{\a\b}-K^{\a\b}\,.
\end{align}
Substituting (\ref{Kon-anh}) and (\ref{CON}) into (\ref{Con-deco}), we
find that the Levi-Civita connection vanishes
\begin{align}
{}^\xi\Gamma^{\a\b}=-K^{\a\b},\qquad
\widetilde{\Gamma}^{\a\b}=0\, .
\end{align}
Moreover, the Riemannian curvature 2-form $\widetilde{R}^{\a\b}$
vanishes due to the trivial Riemannian geometry
\begin{align}
\label{Riemann}
\widetilde{R}^{\a\b}&=\d\, \widetilde{\Gamma}^{\a\b}-
 \widetilde{\Gamma}^{\a\g}\wedge\widetilde{\Gamma}_\g{}^\b=0\, .
\end{align}
Thus, the torsion, the Riemann-Cartan curvature, and the
Riemannian curvature of Cartan's spiral staircase read, respectively,
\begin{align}\label{result}
\boxed{T^\a=2\omega\, \eta^\a\, ,\qquad
R^{\a\b}=\omega^2\, \vt^{\a\b}\, ,\qquad
\widetilde{R}^{\a\b}=0\, .}
\end{align}
This is a very simple configuration. 

In $n$ dimensions, we can decompose the torsion into three
$SO(n)$-irreducible pieces according to (see also~\cite{PRs})
\begin{alignat}{2}
\label{tentor}
^{(1)}T^\a&:=T^\a-\,^{(2)}T^\a-\,^{(3)}T^\a
&&\qquad \text{(tentor)},\\
\label{trator}
^{(2)}T^\a&:=\frac{1}{n-1}\vartheta^\a\wedge\big(e_\b\iner T^\b\big)
&&\qquad\text{(trator)},\\
\label{axitor}
^{(3)}T_\a&:=\frac{1}{3}e_\a\iner\big(\vartheta^\b\wedge T_\b\big)
&&\qquad\text{(axitor)}\, ,
\end{alignat}
with $T^\a={}^{(1)}T^\a+{}^{(2)}T^\a+{}^{(3)}T^\a$. For $n=3$, the
number of independent components is $9=5\oplus 3\oplus 1$. In
(\ref{result}) only the axial torsion part is nonvanishing,
\begin{equation}\label{stairtorsion}
  T^\a=\,^{(3)}T^\a= 2\omega\,\eta^\a\quad\text{or}\quad
{\cal A}:=\frac 13\,^\star\left(\vt^\a\wedge T_\a\right)=2\omega\,.
\end{equation}
In this case the autoparallels of the Riemann-Cartan space coincide
with the Riemannian geodesics (extremals). This is obvious in the
Cartan construction: here the geodesics are just the straight lines of
the underlying Euclidean space---and they are at the same time the
autoparallels in the newly constructed Riemann-Cartan space of
constant axial torsion and constant Riemann-Cartan
curvature.\footnote{Let us stress that Cartan started from a {\it
    flat} 3d Euclidean space, that is, its curvature is {\it zero.} His
  mapping prescription yields the Riemann-Cartan connection
  $^{\xi}\Gamma^{\a\b}$, which is depicted in Figure 1 by means of the
  constantly rotating triads whenever they move in a given
  direction. We find $^{\xi}\Gamma^{\a\b} =
  \omega\,\eta^{\a\b}$. Therefore, our image represents exactly the
  Cartan description. The criticism of Mielke and Maggiolo
  \cite{Mielke:2007} that we ``ignore[s] the constant curvature
  background'' in~\cite{Garcia} is incorrect; after all, a Euclidean
  space carries no nonvanishing curvature.  The original space is
  Euclidean and {\it not} curved. However, the constant curvature
  (\ref{curv}) can be computed from the Riemann-Cartan connection
  $^{\xi}\Gamma^{\a\b}$, which is depicted in our image.}

Of similar simplicity is the Riemann-Cartan curvature. In 3d, the
curvature 2-form $R_\a{}^\b$ is equivalent to the Ricci 1-form
$\text{Ric}_\a:=e_\b\rfloor R_\a{}^\b=\text{Ric}_{i\a}\d x^i$. We have 9
components of the Ricci 1-form. By inspecting (\ref{result}), we
immediately recognize that only the curvature scalar
$R:=e^\a\rfloor\text{Ric}_\a$ is non-vanishing:
\begin{equation}\label{staircurv}
  R^{\a\b}=-\frac 16
  R\,\vt^{\a\b}=\omega^2\,\vt^{\a\b}\quad\text{or}\quad
  R=-6\omega^2\,.
\end{equation}
Thus, Cartan's spiral staircase is characterized geometrically by the
Riemann-Cartan quantities $\left({\cal A}=2\omega,\,R=
  -6\omega^2\right)$ alone, $\cal A$ is a pseudoscalar, $R$ a scalar.

If one decomposes the Riemann-Cartan curvature $R^{\a\b}$ into its
Riemannian part $\widetilde{R}^{\a\b}$ and its rest and then multiplies
with $\eta_{\a\b}$, one finds the geometric identity, see
\cite{PRs,Christian},
\begin{equation}
\label{V2}
  R^{\alpha\beta} \wedge \eta_{\alpha\beta} =
  \widetilde{R}^{\alpha\beta} \wedge \eta_{\alpha\beta} - 2 \, \d
  (\vartheta_\alpha \wedge {}^\star T^\alpha)  +
  T^\alpha \wedge {}^\star \left( - {}^{(1)}T_\alpha + (n-2) \,
    {}^{(2)}T_\alpha + \frac{1}{2} \, {}^{(3)} T_\alpha \right)\,,
\end{equation}
which is valid for any dimension $n$. For Cartan's spiral staircase,
which carries a constant axial torsion, we are left with the 3-form
\begin{equation}
\label{V2ex}
R^{\alpha\beta} \wedge \eta_{\alpha\beta} =
\frac{1}{2}\, {}^{(3)} T^\alpha \wedge {}^\star \, {}^{(3)} T_\alpha \,,
\end{equation}
that is, the curvature is quadratic in the torsion.


\section{The spiral staircase in a Cosserat continuum}
\subsection{Cosserat elasticity\protect\footnote{In
  this subsection we follow in parts the presentation of~\cite{HO07}.}}

The classical continuum of elasticity and fluid dynamics consists of
unstructured points, and the displacement vector $u^\a$ is the only
quantity necessary for specifying the deformation. The Cosserats
conceived a specific {\it medium with microstructure,} see
\cite{Guenther58,Capriz,Erice95,Eringen99} and for a historical review
\cite{Badur}, consisting of structured points such that, in addition
to the displacement field $u^\a$, it is possible to measure the
rotation of such a structured point by the bivector field
$\omega^{\a\b}=-\omega^{\b\a}$, see Figure 2 for a schematic view.

\begin{figure}\label{CossFig1}
\begin{center}
\includegraphics[width=7cm]{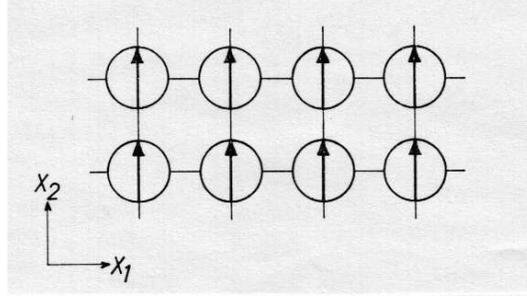}
\end{center}
\caption{Schematic view on a two-dimensional Cosserat continuum:
  Undeformed initial state, see \cite{HO07}.}
\end{figure}

The deformation 1-forms {\it distortion\/} $\beta^\a=\b_i{}^\a\, \d x^i$
and {\it contortion\/} $\kappa^{\a\b}=\kappa_i{}^{\a\b}\, \d x^i$ of a linear
Cosserat continuum are ($\D$ is the exterior covariant derivative of
the Euclidean 3d space)
\begin{eqnarray}\label{Coss1}
  \b^\a&=&\D u^\a + \omega^{\a\b}\vt_\b\,,\qquad\omega^{\a\b}=
  -\omega^{\b\a}\,,\\ \label{Coss2}
  \kappa^{\a\b}&=&\D \omega^{\a\b}\,,
\end{eqnarray}
see G\"unther \cite{Guenther58} and Schaefer \cite{Schaefer}.  For the
components of the distortion, we have
\begin{equation}\label{distortion}
  \b_i{}^\a =\D_i u^\a
  +\omega^{\a\b}e_{i\b} \qquad\text{or}\qquad\b_{\a\b}=\D_\a u_\b 
-\omega_{\a\b}\,.
\end{equation}
In classical elasticity, the only deformation measure is the
strain 
\begin{equation}
\varepsilon_{\a\b}:=\frac 12(\b_{\a\b}+\b_{\b\a})\equiv
\b_{(\a\b)}=\D_{(\a}u_{\b)}\,.
\end{equation} 

Let us visualize these deformations. If the displacement field
$u_1\sim x$ and the rotation field $\omega_{\a\b}=0$, we find
$\b_{11}=\varepsilon_{11}=const$ and $\kappa_{\a\b\g}=0$, see Figure 3. This
homogeneous strain is created by ordinary force stresses. In contrast,
if we put $u_\a=0$ and $\omega_{12}\sim x$, then
$\b_{12}=\omega_{12}\sim x$ and $\kappa_{112}\sim const$, see
Figure 4. This homogeneous contortion is induced by applied spin moment
stresses. Figure 5 depicts the pure constant antisymmetric stress with
$\omega_{12}=const$ and Figure 6 the conventional rotation of the
particles according to ordinary elasticity. This has to be
distinguished carefully {}from the situation in Figure 4.

\begin{figure}\label{CossFig2}
\begin{center}
\includegraphics[width=11cm]{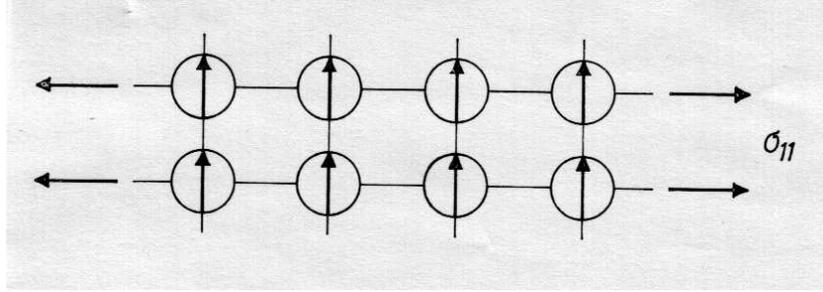}
\end{center}
\caption{Conventional homogeneous strain $\varepsilon_{11}$ of a Cosserat
  continuum: Distance changes of the ``particles'' caused by force
  stress $\sigma_{11}$, see \cite{HO07}.}
\end{figure}

\begin{figure}\label{CossFig3}
\begin{center}
\includegraphics[width=8cm]{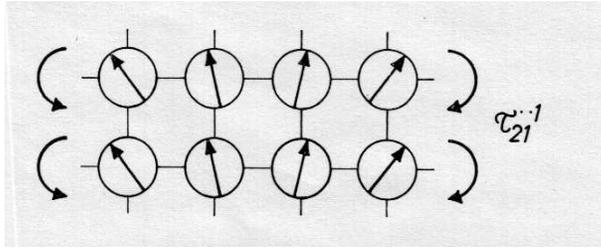}
\end{center}
\caption{Homogeneous contortion $\kappa_{112}$ of a Cosserat continuum:
  Orientation changes of the ``particles'' caused by spin moment
  stress $\tau_{21}{}^1$, see \cite{HO07}.}
\end{figure}

\begin{figure}\label{CossFig4}
\begin{center}
\includegraphics[width=6cm]{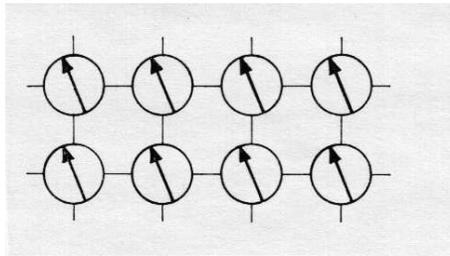}
\end{center}
\caption{Homogeneous Cosserat rotation $\omega_{12}$ of the
  ``particles'' of a Cosserat continuum caused by the antisymmetric
  piece of the stress $\Sigma_{[12]}$, see \cite{HO07}.}
\end{figure}

\begin{figure}\label{CossFig5}
\begin{center}
\includegraphics[width=7cm]{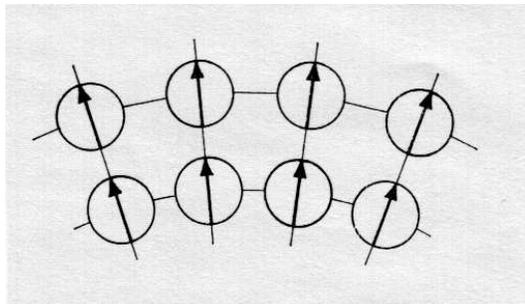}
\end{center}
\caption{Conventional rotation $\partial_{[1}u_{2]}$ of the
  ``particles'' of a Cosserat continuum caused by an inhomogeneous
  strain, see \cite{HO07}.}
\end{figure}

Apparently, in addition to the force stress
$\overline{\Sigma}_{\a}\equiv \Sigma_{\a\b}\eta^\b \sim \pd{\cal
  H}/\pd \b^{\a}$ (here $\cal H$ is an elastic potential), which is
asymmetric in a Cosserat continuum, i.e., ${\Sigma}_{\a\b}\ne
{\Sigma}_{\b\a}$, we have as new response the {\it spin moment} stress
$\overline{\tau}_{\a\b}\equiv\tau_{\a\b}{}^\g\eta_\g\sim\pd{\cal
  H}/\pd \kappa^{\a\b}$. Hence the 2-forms of (force) stress
$\overline{\Sigma}_{\a}$ and of {\it spin moment} stress
$\overline{\tau}_{\a\b}$ characterize a Cosserat continuum {}from the
static side. We used bars for denoting the Cosserat stress and spin
moment stress 2-forms specifically in 3d.

The equilibrium conditions for forces and moments read 
\begin{eqnarray}\label{equilibrium}
  \D\overline{\Sigma}_\a+f_\a=0\,,
  \quad \D\overline{\tau}_{\a\b}+\vt_{[\a}\wedge\overline{\Sigma}_{\b]}
  +m_{\a\b}=0\,.
\end{eqnarray}
where $f_\a$ are the volume forces and $m_{\a\b}=-m_{\b\a}$ volume
moments.  These relations are valid in all dimensions $n\ge 1$, see
\cite{Hyperstress}. In 3 dimensions we have $3+3$ and in 4 dimensions
$4+6$ independent components of the ``equilibrium'' conditions. They
correspond to translational and rotational Noether identities. In
classical elasticity and in fluid dynamics,
$\overline{\tau}_{\a\b}=0$ and $m_{\a\b}=0$; thus, the stress is
symmetric, $\vt_{[\a}\wedge\overline{\Sigma}_{\b]}=0$, and then denoted by
$\overline{\sigma}_{\a}$; for early investigations of asymmetric
stress and energy-momentum tensors, see Costa de Beauregard
\cite{Costa}.

For a local, linear, isotropic Cosserat continuum we have the following
constitutive relations
\begin{align}
\label{CR1-CT}
\overline{\Sigma}_\a &=g_{\a\b}\,^\star\sum_{I=1}^3 \overline{c}_I\,
^{(I)}\beta^{\b}\, ,
\\
\label{CR2-CT}
\overline{\tau}_{\a\b}&=g_{\a\g} g_{\b\de}\,^\star\sum_{I=1}^3 \overline{a}_I\,
^{(I)}\kappa^{\g\de}\, ,
\end{align}
where the 3 irreducible pieces of the distortion 1-form $\beta$ are
given by
\begin{eqnarray}
  \beta^\a=\,^{(1)}\beta^\a +
  \,^{(2)}\beta^\a + \,^{(3)}\beta^\a\,,
\end{eqnarray}
with the number of independent
components $9=5\oplus 3\oplus 1$. In component notation, they are given by
\begin{eqnarray}
  ^{(1)}\beta_{\a\b}&:=&\beta_{\a\b}-\,^{(2)}\beta_{\a\b}-\,^{(3)}\beta_{\a\b}\,,\\
^{(2)}\beta_{\a\b}&:=& \frac{1}{2}\,\big(\beta_{\a\b}-\beta_{\b\a}\big) \, ,\\
  ^{(3)}\beta_{\a\b}&:=& \frac{1}{3}\, \delta_{\a\b} \, \beta_\g{}^\g \,.
\end{eqnarray}
In addition, the 3 irreducible pieces of the contortion 1-form
$\kappa$ are given by
\begin{eqnarray}
  \kappa^{\a\b}=\,^{(1)}\kappa^{\a\b} +
  \,^{(2)}\kappa^{\a\b} + \,^{(3)}\kappa^{\a\b}\,,
\end{eqnarray}
with the number of independent
pieces $9=5\oplus 3\oplus 1$. In component notation, they are given by
\begin{eqnarray}
  ^{(1)}\kappa_{\a\b\g}&:=&\kappa_{\a\b\g}-\,^{(2)}\kappa_{\a\b\g}
  -\,^{(3)}\kappa_{\a\b\g}\,,\\
  ^{(2)}\kappa_{\a\b\g}&:=& \frac{1}{2}\,\big(\delta_{\a\b}\,
  \kappa^\de{}_{\de\g}-\delta_{\a\g}\kappa^\de{}_{ \de\b}\big) \, ,\\
  ^{(3)}\kappa_{\a\b\g}&:=& \frac{1}{3}\, \big(
  \kappa_{\a\b\g}+\kappa_{\b\g\a}+\kappa_{\g\a\b}\big)\,.
\end{eqnarray}
The nonnegative moduli $\overline{c}_1$, $\overline{c}_2$,
$\overline{c}_3$ have the dimension:
$[\overline{c}_I]=\text{force}/(\text{length})^2$ and
$\overline{a}_1$, $\overline{a}_2$, $\overline{a}_3$ have the
dimension: $[\overline{a}_I]=\text{force}$.  In this way, the elastic
potential for local, linear and isotropic Cosserat theory reads
\begin{align} 
{\cal H}=\frac{1}{2}\, \overline{\Sigma}_\a\wedge\b^\a+ 
\frac{1}{2}\, 
  \overline{\tau}_{\a\b}\wedge \kappa^{\a\b}\,.
\end{align}

Nowadays the Cosserat continuum finds many applications. As one
example we may mention the work of Zeghadi {\it et al.} \cite{grains} who
take the grains of a metallic polycrystal as (structured) Cosserat
particles and develop a linear Cosserat theory with the constitutive
laws $\overline{\Sigma}_{\a}\sim \b_{\a}$ and
$\overline{\tau}_{\a\b}\sim \kappa_{\a\b}$.

The Riemannian space is the analogue of the body of classical
continuum theory: points and their relative distances is all what is
needed to describe it geometrically; the analogue of the strain
$\varepsilon_{ij}$ of classical elasticity is the difference between
the metric tensor $g_{ij}$ of the Riemannian space and a flat
background metric. In GR, a symmetric ``stress''
$\sigma_{ij}=\sigma_{ji}$ is the response of the matter Lagrangian to
a variation of the metric $g_{ij}$.

A RC-space can be realized by a generalized Cosserat continuum.  The
``deformation measures'' $\vt^\a=e_i{}^\a\, \d x^i$ and
$\Gamma^{\a\b}=\Gamma_i{}^{\a\b}\d x^i=-\Gamma^{\b\a}$ of a RC-space
correspond to those of a Cosserat continuum according to the
transcription\footnote{This can be seen {}from the response of the
  coframe $e_i{}^\a$ and the Lorentz connection $\Gamma_i{}^{\a\b}$ in
  a RC-space to a local Poincar\'e gauge transformation consisting of
  small translations $\epsilon^\a$ and small Lorentz transformations
  $\omega^{\a\b}$,
\begin{eqnarray}\label{PG1}
  \d e_i{}^\a &=&\, -\D_i\e^\a+e_i{}^\g\omega_\g{}^\a-\e^\g T_{\g i}{}^\a\,,\\
  \d\Gamma_i{}^{\a\b}&=&\,   -\D_i\omega^{\a\b}
-\e^\g R_{\g i}{}^{\a\b}\,,\label{PG2}
\end{eqnarray}
see \cite{RMP}, Eqs.(4.33),(4.32); here $\D_i:=\partial_i\rfloor \D$ are
the components of the exterior covariant derivative.  The second term
on the right-hand-side of (\ref{PG1}) is due to the semi-direct
product structure of the Poincar\'e group. If we put torsion and
curvature to zero, these formulas are analogous to
(\ref{Coss1}),(\ref{Coss2}).}
\begin{equation}\label{map}
  \delta e_i{}^\a\rightarrow \b^\a\,,\qquad\delta\Gamma_i{}^{\a\b}
  \rightarrow \kappa^{\a\b}\,.
\end{equation}
However, in general, the coframe $\vt^\a$ and the connection $\Gamma^{\a\b}$
cannot be derived {}from a displacement field $u^\a$ and a rotation field
$\omega^{\a\b}$, as in (\ref{Coss1}) and (\ref{Coss2}). Such a generalized
Cosserat continuum is called incompatible, since the deformation
measures $\b^{\a}$ and $\kappa^{\a\b}$ don't fulfill the so-called
compatibility conditions
\begin{eqnarray}\label{compat}
\D\b^\a+\vt^\b\wedge\kappa_\b{}^\a=0\,,\qquad \D\kappa^{\a\b}=0\,,
\end{eqnarray}
see G\"unther \cite{Guenther58} and Schaefer
\cite{SchaeferZAMM,Schaefer}. They guarantee that the ``potentials''
$u^\a$ and $\omega^{\a\b}$ can be introduced in the way as it is done in
(\ref{Coss1}) and (\ref{Coss2}). Still, also in the RC-space, as {\it
  incompatible\/} Cosserat continuum, we have, besides the force
stress $\overline{\Sigma}_\a{}\sim\pd{\cal H}/\pd \b^\a$, the
spin moment stress $\overline{\tau}_{\a\b}{}\sim\pd{\cal H}/\pd
\Gamma^{\a\b}$.  And in the geometro-physical interpretation of
the structures of the RC-space, Cartan apparently made use of these
results of the brothers Cosserat.

In 4d, the stress $\overline{\Sigma}_\a$ corresponds to
energy-momentum\footnote{This is well-known {}from classical
  electrodynamics: The 3d Maxwell stress generalizes, in 4d, to the
  energy-momentum tensor of the electromagnetic field, see
  \cite{Birkbook}.} ${\Sigma}_\a$ and the spin moment stress
$\overline{\tau}_{\a\b}$ to spin angular momentum
${\tau}_{\a\b}$. Accordingly, Cartan enriched the 4d Riemannian space
of GR geometrically by the {\it torsion} 2-form $T^\a$ and statically
(or dynamically) by the {\it spin angular momentum} 3-form $\tau_{\a\b}$
of matter.

\subsection{Incompatible Cosserat elasticity}

In order to include the torsion tensor and the curvature tensor into the
framework of Cosserat elasticity, we have to generalize compatible
Cosserat elasticity to incompatible Cosserat elasticity.  Such an
extension is necessary for Cartan's spiral staircase in the framework
of Cosserat theory since we have already seen that the Cartan spiral
staircase is related to the notion of torsion~(\ref{torsion}).

In the case of incompatible Cosserat elasticity,
the distortion and the contortion do not satisfy any longer the compatibility 
conditions~(\ref{compat}).
Then the elastic distortion and the elastic contortion are given as
\begin{align}
\label{dist-inc}
  \b^\a&=\D u^\a + \omega^{\a\b}\vt_\b-{}^{{\rm P}}\!\b^\a\,,\\
  \kappa^{\a\b}&=\D \omega^{\a\b}-{}^{\rm P}\!\kappa^{\a\b}\,.
\label{cont-inc}
\end{align}
It can be seen that the plastic distortion $^{\text{P}}\!\beta^\a$ and
the plastic contortion $^{\text{P}}\!\kappa^{\a\b}$ are the causes of the
incompatibility.  The failure of the elastic and plastic fields to be
compatible gives rise to incompatibility
conditions~\citep{Guenther58,Schaefer}:
\begin{align}
\label{IC1-me}
T^\a&=\D\b^\a+\kappa^{\b\a}\wedge \vt_\b\,,\\
\label{IC2a-me}
R^{\a\b}&= \D\kappa^{\a\b}\,,
\end{align}
and for the plastic fields
\begin{align}
\label{IC1-p-me}
T^\a&=-\D\, {}^{\text{P}}\!\b^\a-{}^{\text{P}}\!\kappa^{\b\a}\wedge \vt_\b\,,\\
R^{\a\b}&= -\D\, {}^{\text{P}}\!\kappa^{\a\b}\,.
\label{IC2a-p-me}
\end{align}
The measures of incompatibilities~(\ref{IC1-me})--(\ref{IC2a-p-me})
may be identified with the torsion 2-form and the curvature 2-form of
the incompatible Cosserat medium in linear approximation.

\subsection{Cartan's spiral staircase as a solution in incompatible 
Cosserat elasticity}
In this subsection we want to show, that the solutions~(\ref{conncomp2}) and 
(\ref{coframe}) of Cartan's spiral staircase 
are also solutions in incompatible Cosserat elasticity. 
If we use the identification~(\ref{map}), we find
for the elastic distortion and the elastic contortion
\begin{align}
\label{dist-Cartan-CT}
\beta_{\a\b}=\delta_{\a\b}\, ,\qquad
\kappa_{\a\b\g}=\omega\, \eta_{\a\b\g}\, .
\end{align}
Thus, the elastic distortion and the elastic contortion are constant.
The irreducible pieces are 
\begin{alignat}{3}
\label{B-Ir}
&^{(1)}\beta_{\a\b}=0\, ,\qquad 
&^{(2)}\beta_{\a\b}= 0 \,,\qquad
&^{(3)}\beta_{\a\b}=\delta_{\a\b} \, ,\\
\label{cont-Ir}
&^{(1)}\kappa_{\a\b\g}=0\, ,\qquad 
&^{(2)}\kappa_{\a\b\g}= 0 \,,\qquad
&^{(3)}\kappa_{\a\b\g}=\omega\, \eta_{\a\b\g} \,.
\end{alignat}
If we substitute (\ref{dist-Cartan-CT}) into (\ref{dist-inc}) and
(\ref{cont-inc}) and integrate, we find for the displacement and the
microrotation bivector
\begin{align}
u_\a =x_\a\, ,\qquad \omega_{\a\b}=\omega\, \eta_{\a\b\g} x^\g
\end{align}
and for the plastic distortion and plastic contortion
\begin{align}
  ^{\text{P}}\!\beta_{\a\b}=-\omega_{\a\b}\, \qquad
  ^{\text{P}}\!\kappa_{\a\b\g}=0\, .
\end{align}
Thus, the plastic rotation is equal to the negative microrotation
bivector and the plastic contortion is zero.  From
(\ref{IC1-me})--(\ref{IC2a-p-me}) we calculate the torsion and the
curvature produced by Cartan's spiral staircase in linear,
incompatible Cosserat elasticity as
\begin{align}
T^\a=2\omega\, \eta^\a\, ,\qquad R^{\a\b}=0\, .
\end{align}
The vanishing of $R^{\a\b}$ identifies the corresponding RC-space as a
teleparallel one.

Now we may substitute (\ref{B-Ir}) and (\ref{cont-Ir}) into the
constitutive relations (\ref{CR1-CT}) and (\ref{CR2-CT}) and we find
for the force and internal moment stresses caused by Cartan's spiral
staircase
\begin{align}
  \label{Stress-CT} {\Sigma}_{\a\b}=-p\,
  \delta_{\a\b}=\overline{c}_3\,\delta_{\a\b}\, ,\qquad
  {\tau}_{\a\b\g}=\overline{a}_3 \omega\, \eta_{\a\b\g}\, .
\end{align}
Mechanically, 
we have found a constant hydrostatic {\it pressure} $-\overline{c}_3$ and
a constant {\it torque}\,  $\overline{a}_3 \omega$ as predicted by Cartan.
Thus, we conclude that Cartan's spiral staircase is a solution in linear,
 incompatible Cosserat theory producing constant pressure and constant internal 
torque in a Cosserat medium.

\section{The spiral staircase in three-dimensional theories of
  gravity}

In the realm of quantum gravity, people are interested in
(1+2)-dimensional theories of gravity, basically since
(1+3)-dimensional theories, like GR or the Einstein-Cartan theory, in
some high `temperature' limit, may effectively reduce to
(1+2)-dimensional theories. A good reference describing this approach
is Carlip \cite{Carlip:1998}. We concentrate here purely on the
classical aspect of these theories.

The conventional gravitational Lagrangian in 4d is the
Hilbert-Einstein Lagrangian $\sim \eta_{\a\b}\wedge R^{\a\b}$. This
term also works in 3d. However, in 3d there exist topological,
connection-dependent terms, namely the Chern-Simons 3-form for the
curvature
\begin{equation}\label{ChernSimons}
  C_{\text{RR}} :=-\frac{1}{2}\left(\Gamma_\a{}^\b\wedge \d\Gamma_\b{}^\a
    -\frac{2}{3}\,\Gamma_\a{}^\b\wedge\Gamma_\b{}^\gamma\wedge
\Gamma_\gamma{}^\a\right)\,.
\end{equation}
This equation is correct in a Riemann or a Riemann-Cartan space, for
details, see, e.g., \cite{PRs}, Sec.3.9. In a Riemann-Cartan space we
can define an analogous 3-form for the torsion, namely
\begin{equation}\label{CStorsion}
C_{\text{TT}}:=\frac{1}{2\ell^2}\,g_{\a\b}\,\vt^\a\wedge T^\b\,,
\end{equation}
where $\ell$ is some constant with the dimension of a
length. Introducing additionally a cosmological term with $\Lambda$ as
cosmological constant, we end up with the Mielke-Baekler Lagrangian
\cite{MB91,BMH92}, see also \cite{Mielke:2007},
\begin{eqnarray}\label{MB}
  V_{\text{MB}}&=&-\frac{\chi}{2\ell}\,R^{\a\b}\wedge\eta_{\a\b}
  -\frac{\Lambda}{\ell}\,\eta+\frac{\theta_{\rm T}}{2\ell^2}\,\vt^\a\wedge T_\a
  \nonumber\\ 
  &&-\frac{\theta_{\rm L}}{2}\left(\Gamma_\a{}^\b\wedge \d\Gamma_\b{}^\a
    -\frac{2}{3}\,\Gamma_\a{}^\b\wedge\Gamma_\b{}^\gamma\wedge
    \Gamma_\gamma{}^\a\right)+L_{\text{mat}}\,,
\end{eqnarray}
with some coupling constants $\chi,\theta_{\rm T}, \theta_{\rm L}$ (here `T' stands for translation and `L' for Lorentz). Theories with this general Lagrangian will be considered.

\subsection{3d Einstein-Cartan theory}

In the simplest case we just have, for $\chi=1$, the 3d
Einstein-Cartan Lagrangian without cosmological constant,
\begin{equation}\label{EC}
V_{\text{EC}}=-\frac{1}{2\ell}R^{\a\b}\wedge\eta_{\a\b}+L_{\text{mat}}\,.
\end{equation}
Variations with respect to coframe $\vt^\a$ and Lorentz connection
$\Gamma^{\a\b}$, yield the field equations of the 3d Einstein-Cartan
theory (with Euclidean signature) \cite{PRs,Garcia,Trautman}:
\begin{align}
\label{EC-fe1}
\frac{1}{2}\, \eta_{\a\b\g}\, R^{\b\g}&=\ell\, \Sigma_{\a}\,,\\
\label{EC-fe2}
\frac{1}{2}\, \eta_{\a\b\g}\, T^{\g}&=\ell\, \tau_{\a\b}\,,
\end{align}
where $\Sigma_\a$ and $\tau_{\a\b}$ are the 2-forms of (force) stress
and of (spin) moment stress, respectively. Moreover, $\ell$ is a
characteristic length.\footnote{Roughly speaking, we could imagine
  $\ell$ as the distance between neighboring dislocation lines of the
  dislocation forests mentioned above, see \cite{HK}; depending on the
  state of the crystal, this length $\ell$ could typically be of the
  order of some $50$ nm.}

Substituting~(\ref{curv}) and (\ref{torsion}) into (\ref{EC-fe1}) and
(\ref{EC-fe2}), respectively, and using simple algebra, we find the
force stress 2-form and the moment stress 2-form,\footnote{Also here
  the signs in \cite{HO07} are opposite.}
\begin{align}\label{qaz}
  \Sigma_\a=\frac{\omega^2}{\ell}\,
  \eta_\a\,,\qquad \tau_{\a\b} =\frac{\omega}{\ell}\,
  \vt_{\a\b}\, .
\end{align}
In order to find the tensor components, we develop the 2-forms
$\Sigma_\a$ and $\tau_{\a\b}$ with respect to the 2-form $\eta_\a$:
\begin{align}\label{qazq}
  \Sigma_\a=:{\mathfrak t}_\a{}^\b\,\eta_\b\,,\qquad \tau_{\a\b}
  =:{\mathfrak s}_{\alpha\beta}{}^\gamma\,\eta_\gamma\, .
\end{align}
Inversion of (\ref{qazq}) and use of (\ref{qaz}) yields for the force
stress tensor and the moment stress tensor
\begin{align}
\label{stress-EC}
{\mathfrak t}_\alpha{}^\beta= -p\de_\a^\b=\frac{\omega^2}{\ell} \,
\delta_\alpha^\beta\,,\qquad
{\mathfrak s}_{\alpha\beta\gamma}=\frac{\omega}{\ell} \,\eta_{\alpha \beta
    \gamma}\,.
\end{align}
We have found a constant hydrostatic {\it pressure} $ -\omega^2/\ell$
and a constant {\it torque} $\omega/\ell$, exactly as foreseen by
Cartan.
Thus, the spiral staircase is an exact solution of the 3d Einstein-Cartan
theory (with Euclidean signature) carrying constant pressure and constant
torque as sources.

\subsection{3d Poincar\'e gauge theory of gravity with Mielke-Baekler
  Lagrangian}

The EC-theory has a Lagrangian linear in the curvature. As a
consequence, the Lorentz connection is non-propagating. If one allows
for higher order terms, as in the Mielke-Baekler Lagrangian, the
Lorentz connection becomes `liberated'. Such theories are Poincar\'e
gauge theories. By variation of the Mielke-Baekler Lagrangian one
arrives at the field equations
\begin{eqnarray}\label{MBfieldeq}
  \frac{\chi}{2}\,\eta_{\a\b\g}\,R^{\b\g}+\Lambda\,\eta_\a-\frac{\theta_{\rm T}}{\ell}\,T_\a&=&\ell\,\Sigma_\a\,,\\
  \frac{\chi}{2}\,\eta_{\a\b\g}\,T^\g-\frac{\theta_{\rm T}}{2\ell}\,\vt_{\a\b}-\theta_{\rm L}\ell\,R_{\a\b}&=&\ell\, \tau_{\a\b}\,.
\end{eqnarray}

Garc\'{\i}a {\it et al.}\ \cite{Garcia} looked for static circularly
symmetric {\it vacuum} solutions of these field equation. In fact, for
the 3d Einsteinian case in a Riemannian space such a solution had been
found by Ba{\~n}ados {\it et al.} \cite{BTZ}. Garc\'{\i}a {\it et al.} generalized
this so-called BTZ-solution (Ba{\~n}ados, Teitelboim, Zanelli) to a
`BTZ-solution with torsion' \cite{Garcia}. The details for this
solution can be found in \cite{Garcia}, Table I. If one puts the
effective cosmological constant to zero, $\Lambda_{\rm eff}=0$, this
vacuum solution has the torsion and curvature
\begin{equation}\label{BTZ}
  T^\a =2\frac{\cal T}{\ell}\,\eta^\a\,,\qquad R^{\a\b}=-\left(\frac{{\cal T}}{\ell}\right)^2\vt^{\a\b}\,,\qquad\widetilde{R}^{\a\b}=0\,.
\end{equation}
Here the constant ${\cal T}$ can be expressed in terms of the coupling
constants according to
\begin{equation}
{\cal T}=-\frac{\theta_{\rm T}\chi}{2(\chi^2+2\theta_{\rm T})\theta_{\rm L}}\,.
\end{equation}
If we compare (\ref{BTZ}) with (\ref{result}), we see that (apart from
a probably signature dependent sign) the torsion and the curvatures
coincide. Consequently, a subcase of the vacuum BTZ-solution with
torsion carries the torsion and the curvature of Cartan's spiral
staircase. We stress that, in contrast to the solution (\ref{EC-fe1})
with (\ref{EC-fe2}), where we only have to assume constant sources, in
(\ref{BTZ}) we have an exact {\it vacuum} solution. This was outside
the scope of Cartan in 1922.

\section{The translation gauge theory of dislocations in three
  dimensions}

Let be given a solid body with crystalline structure. Often such
solids contain lattice defects that may be created during the growing
of the crystal or during plastic deformation. One-dimensional lattice
defects are dislocation lines that (in a cubic primitive crystal) are
of two types: Edge dislocations (see Figure 7) and screw dislocations
(see Figure 8). From comparing Figure 8 with the spiral staircase
Figure 1, it is clear that the geometry of Cartan's spiral staircase can be
represented by a set of three perpendicular constant `forests' of {\it
  screw dislocations} of equal strength.  It is our goal to show that
the spiral staircase emerges as a solution in the framework of the
gauge theory of dislocations. In this theory the real crystal,
containing dislocations, is modeled as a 3-dimensional space with
teleparallelism (Weitzenb\"ock space) the torsion of which represents
the dislocation density.  By a suitable choice of the frames, it is
always possible to `gauge' the connection 1-form of the Weitzenb\"ock
space globally to zero \cite{Nester:2010},
\begin{equation}
\label{conn0}
\Gamma^{\a\b}=0\qquad\text{(in suitable frames)}\,.
\end{equation}
We dropped here the $\xi$ that designated in Section 2 the
non-Riemannian connection.
\begin{figure}\label{dislFig1}
\begin{center}
\includegraphics[width=7cm]{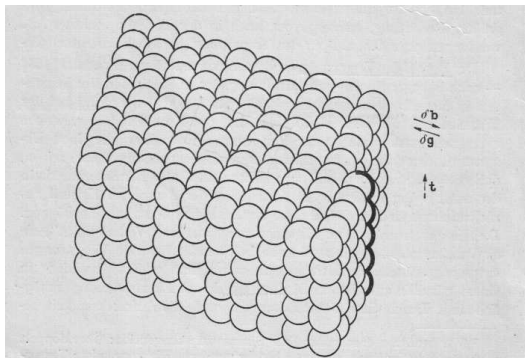}
\end{center}
\caption{{\it Edge dislocation} after Kr\"oner \cite{Kroener1958}: The
  dislocation line is parallel to the vector $\mathbf{t}$. The Burgers
  vector $\delta\mathbf{b}$, characterizing the missing half-plane, is
  perpendicular to $\mathbf{t}$. The vector $\delta\mathbf{g}$
  characterizes the gliding of the dislocation as it enters the ideal
  crystal.}
\end{figure}
\begin{figure}\label{dislFig2}
\begin{center}
\includegraphics [width=7cm]{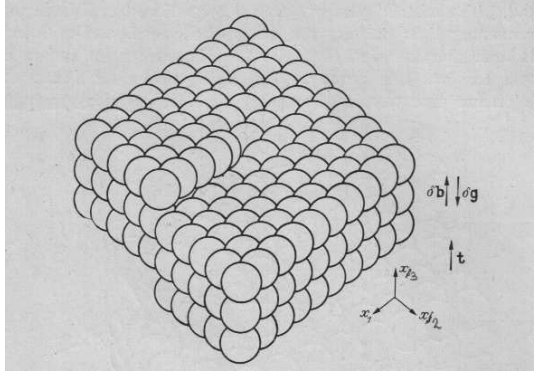}
\end{center}
\caption{{\it Screw dislocation} after Kr\"oner \cite{Kroener1958}: Here
the Burgers vector is parallel to $\mathbf{t}$.}
\end{figure}

\subsection{Foundations}

Let us display the structure of the three-dimensional translational
gauge theory of dislocations as proposed by
Lazar~\cite{Lazar00,Lazar02}. This theory follows the basic features
of the metric-affine gauge theory of gravity, see the review
\cite{PRs}.

At the outset we identify the torsion 2-form
$T^\a$ with the {\it dislocation density}. Since subsequently we always
assume orthonormal frames that nullify the
connection according to (\ref{conn0}), the torsion 2-form reads
\begin{align}
\label{Torsion}
T^\a=\d\vartheta^\a=\frac{1}{2}\, T_{\b\g}{}^\a \,\vt^\b\wedge\vt^\g.
\end{align}
Accordingly, the torsion $T^\a$, as long as the gauge condition
(\ref{conn0}) is fulfilled, corresponds to the object of anholonomity
$\Omega^\a:=\d\vt^\a$.  

For the physical interpretation of quantities, the knowledge of the
physical dimensions is decisive. The dimension of the coframe $\vt^\a$
is $[\vt^\a]=\text{length}$ and $[e_\a]=1/\text{length}$. Thus, the
dimension of $T^\a$ is $[T^\a]=\text{length}$---it is called the {\it
  absolute} dimension of $T$--- and the {\it physical} dimension (of
its components) turns out to be $[T_{\b\g}{}^\a]=1/\text{length}$.
The torsion 2-form satisfies the first (or translational) Bianchi
identity
\begin{align}
\label{bianchi}
\d T^\a=0\,.
\end{align}

In a second step, the coframe 1-form $\vt^\a$ is identified with the
(incompatible) {\it elastic distortion} 1-form known from continuum
mechanics. We decompose the distortion 1-form into its components,
\begin{align}
\label{dist1}
\vt^\a=e_i{}^\a\, \d x^i\, .
\end{align}
The physical dimension of them is $[e_i{}^\a]=1$.  Accordingly, we
take $\vt^\a$ and $T^\a$ as field variables of the dislocation gauge
theory.

Now we can set up the total Lagrangian 3-form $L_{\text{tot}}$
describing dislocations in an incompatible elastic continuum, with
absolute dimension $[L_{\text{tot}}]=\text{energy}$. It is given by
the sum of the {\it elastic} Lagrangian $L$ of the material continuum
and the {\it gauge} Lagrangian $V$ of the dislocation fields:
\begin{align}
L_{\text{tot}}=L(\vt^\a)+V(\vt^\a,T^\a)\,.
\end{align}
The covector-valued 2-form of the elastic (force) stress is defined by
\begin{align}
\label{stress_el}
\Sigma^{}_\a:=\frac{\delta L}{\delta \vartheta^\a}\,.
\end{align}
It has the absolute dimension $[\Sigma_\a]=\text{force}$.
In general, this stress $\Sigma_\a$ is asymmetric.  Analogously to
(\ref{stress_el}), we can define the dislocation stress, a
covector-valued 2-form, as
\begin{align}
\label{E}
E_\a:=\frac{\pd V}{\pd\vartheta^\a}
=e_\a\iner V+\big(e_\a\iner T^\b\big)\wedge H_\b\,.
\end{align}
the absolute dimension of which is $[E_\a]=\text{force}$.  Eshelby
\cite{Eshelby} called such a type of expression an ``elastic
energy-momentum''; if $E_\a$ is integrated over a 2-dimensional closed
surface, it yields the force on defects (here dislocations) within the
surface. Similar to $\Sigma_\a$, the dislocation stress $E_\a$ is
asymmetric in general.

Torsion $T^\a$ has the status of a gauge field strength in the formalism.
Accordingly, we can define the attached {\it excitation} 1-form as
\begin{align}
\label{H}
H_\a:=-\frac{\pd V}{\pd T^\a}\, ,
\end{align}
with $[H_\a]=\text{force}$. It is the response of the gauge Lagrangian
$V$ to the torsion 2-form $T^\a$.

The moment stress $\tau_{\a\b}=-\tau_{\b\a}$ is coupled to the contortion 1-form
$K_{\a\b}=-K_{\b\a}$ according to
\begin{equation}\label{momstress}
\tau_{\a\b}=\frac{\pd V}{\pd K^{\a\b}}\,,\qquad\text{and}\qquad
\tau_{\a\b}=\vt_{[\a}\wedge H_{\b]}\,,
\end{equation}
with the dimension
$[\tau_{\a\b}]=\text{force}\times\text{length}=\text{moment}$, since
$[K_{\a\b}]=1$.  The last formula in (\ref{momstress}), analogously to
(\ref{contortion1}), can be inverted as follows:
\begin{equation}\label{last}
  H_\alpha=-2e_\beta\rfloor \tau_\alpha{}^\beta + \frac{1}{2}\,
  \vartheta_\alpha\wedge(e_\beta\rfloor e_\gamma\rfloor\tau^{\beta\gamma})\,\,.
\end{equation}

Since, according to (\ref{Torsion}), $T^\a$ depends on $\vt^\a$, the
independent variable of the variational principle is
$\vt^\a$. Independent variation yields the {\it Euler-Lagrange
  equation} of dislocation gauge theory:
\begin{align}
\label{EL}
\frac{\delta L_{\text{tot}}}{\delta \vt^\a}\equiv \d\,\frac{\pd
  L_{\text{tot}}}{\pd T^\a} +\frac{\pd L_{\text{tot}}}{\pd
  \vt^\a}=0\,.
\end{align}
By means of (\ref{H}), (\ref{stress_el}), and (\ref{E}), we can
rewrite it as:
\begin{align}
\label{YM-fe}
\d H_\a-E_\a=\Sigma^{\rm }_\a\,.
\end{align}
This is a Yang-Mills type field equation. The sum of the two stresses
$\Sigma_\a$ and $E_\a$ constitutes the source of the excitation $H_\a$.

The field equation implies the {\it force equilibrium}: we
differentiate (\ref{YM-fe}) and find the law
\begin{align}\label{forceequi}
\d (E_\a+ \Sigma_\a)=0\, .
\end{align}
The total stress is apparently in equilibrium. From this equation we
can read off the covector-valued {\it Peach-Koehler force} 3-form as
\begin{align}
\label{PKF}
  f_\alpha:=\d \Sigma_\a= -\d E_\a= (e_\a \iner T^\b)\wedge
  \Sigma_\b\,.
\end{align}
It represents the force density acting on a dislocation.

The {\it moment equilibrium} requires a bit of algebra. We start with
the field equation (\ref{YM-fe}) and compute the antisymmetric piece
of the total stress, use (\ref{momstress}), and find
\begin{equation}\label{momequi}
\d\tau_{\a\b}-T_{[\a}\wedge H_{\b]}+\vt_{[\a}\wedge\left( E_{\b]}
    +\Sigma_{\b]}\right)=0\,.
\end{equation}
Apart from the nonlinear term $-T_{[\a}\wedge H_{\b]}$, this is
exactly the expected law.

This represents the general framework of the dislocation gauge
theory. Now we have to specify a {\it constitutive laws.} For the {\it
  excitation} in a local, linear, isotropic continuum we have
\begin{align}
\label{const_iso}
H_\a=\,\hodge\!\sum_{I=1}^{3}a_{I}\,^{(I)}T_\a\, ,
\end{align}
wherein $^{(I)}T_\a$ are the irreducible pieces (\ref{tentor}),
(\ref{trator}), and (\ref{axitor}) of the torsion and $a_1$, $a_2$,
and $a_3$ nonnegative constitutive moduli with dimension:
$[a_I]=\text{force}$.  For the elastic (force) {\it stress}, we assume
a Hooke type law
\begin{align}
\label{const_Sigma}
\Sigma_\a= \,^\star\sum_{I=1}^{3} c_I \,^{(I)}\vt_\a \,,
\end{align}
where $c_1$, $c_2$, and $c_3$ are nonnegative constitutive moduli with
the dimension: $[c_I]=\text{force}/(\text{length})^2$.  Here
$^{(I)}\vt_\a$ denotes the irreducible pieces of the elastic
distortion 1-form.  

We want to mention a technical difficulty in the decomposition of
$\vt^\a$.  In the decomposition of the torsion into the irreducible
pieces (\ref{tentor}), (\ref{trator}), and (\ref{axitor}) we used the
coframe and the frame for the exterior and interior multiplication
operations. However, for the decomposition of the coframe itself the
same procedure does not work. Therefore, we have to use different
coframe and frame fields as reference standards for the decomposition
of $\vt^\a$. For simplicity, we choose the specific holonomic coframe
$\hat{\vt}^\a=\delta^\a_i\d x^i$ and the corresponding frame
$\hat{e}_\a=\delta_\a^i \pd_i$. Then, the irreducible pieces are defined
by
\begin{align}
\label{vt1}
^{(1)}\vt^\a&:=\vt^\a-\,^{(2)}\vt^\a-\,^{(3)}\vt^\a, \\
\label{vt2}
^{(2)}\vt_\a&:=\frac{1}{2}\,\hat e_\a\iner\big(\hat\vt_\b\wedge \vt^\b\big),\\
\label{vt3}
^{(3)}\vt^\a&:=\frac{1}{3}\, \hat\vartheta^\a \big(\hat e_\b\iner \vt^\b\big),
\end{align}
with $\vt^\a={}^{(1)}\vt^\a+{}^{(2)}\vt^\a+{}^{(3)}\vt^\a$. The
number of independent components is $9=5\oplus 3\oplus 1$.
In components, the irreducible pieces read
\begin{align}
  ^{(1)}\vt^\a&=\vt^\a-\,^{(2)}\vt^\a-
  \,^{(3)}\vt^\a,\\
  ^{(2)}\vt^\a&=
  \frac{1}{2}\,\big(e_i{}^\a-\delta_{i\b}\delta^{j\alpha}e_j{}^\b\big)
  \d x^i = \frac{1}{2}\,\big(e_i{}^\a-e^\a _{\ \, i}\big) \d
  x^i,\\
  ^{(3)}\vt^\a&=\frac{1}{3}\, \delta^\a_i\delta^j_\b \, e_j{}^\b \, \d
  x^i=\frac{1}{3}\, \delta^\a_i\, e_\b{}^\b \, \d x^i\, .
\end{align}

Let us now return to (\ref{const_iso}) and (\ref{const_Sigma}). Using
these equations and Euler's theorem for homogeneous functions, we can
rewrite the gauge (or dislocation) Lagrangian and the elastic
Lagrangian, respectively, as
\begin{align}
\label{V-2}
V=-\frac{1}{2}\, H_\a\wedge T^\a= \frac{1}{2}\,\tau_{\a\b} \wedge
K^{\a\b}\quad\text{and}\quad L=\frac{1}{2}\, \Sigma_\a\wedge\vt^\a\,.
\end{align}
Summing up, the dislocation theory we displayed encompasses two kinds
of asymmetric force stresses (namely $E_\a$ and $\Sigma_\a$) and one
type of moment stress $\tau_{\a\b}$. Equivalent to $\tau_{\a\b}$ is
the excitation $H_\a$ that plays a fundamental role in the Yang-Mills
type field equation (\ref{YM-fe}). Compactly written, we have
\begin{eqnarray}\label{fund1''}
 T^\a&=&\d\vt^\a\,,\\
\label{fund2''}\d H_\a-E_\a&=&\Sigma_\a\,,\\
\label{fund3''}H_\a&:=&-\frac{\pd {V}}{\pd T^\a}\qquad\text{or}\qquad
H_\a\approx g_{\a\b}
\,^\star\sum_{I=1}^{3} a_I \,^{(I)}T^\b\,,\\
\label{fund4''}E_\a&:=&\frac{\pd {V}}{\pd\vt^\a}
=e_\a\iner {V}+\big(e_\a\iner T^\b\big)\wedge H_\b\,,\\
\label{fund5''}\Sigma^{}_\a&:=&\frac{\delta L}{\delta \vt^\a}
\quad\qquad\text{or}\qquad\Sigma_\a\approx g_{\a\b}\,^\star\sum_{I=1}^{3} c_I
\,^{(I)}\vt^\b \,,\\
\label{fund6''} f_\alpha&=& (e_\a \iner T^\b)\wedge
\Sigma_\b\,
\end{eqnarray}
\noindent($\vt^\a$ = distortion, $T^\a$ = torsion = dislocation
density, $V$ = gauge Lagrangian $\sim$ torsion$^2$, $H_\a$ =
excitation, $\Sigma_\a$ = force stress, $E_\a$ = dislocation stress,
$L$ = matter Lagrangian $\sim$ distortion$^2$, $f_\a$ = Peach-Koehler
force density).

We could add simple consequences of the scheme, namely the homogeneous
field equation and an alternative version of the inhomogeneous field
equation,
\begin{eqnarray}
 \d T^\a&=&0\,,\\
 \d\tau_{\a\b} -T_{[\a}\wedge H_{\b]}+\vt_{[\a}\wedge
  E_{\b]}&=&\vt_{[\a}\wedge
  \Sigma_{\b]}\,,
\end{eqnarray}
with the spin moment stress $\tau_{\a\b}=\vt_{[\a}\wedge H_{\b]}$.


In a more recent development, one of us ref.~\cite{L09} investigated
the Higgs mechanism in the framework of the translation gauge theory
of dislocations. At the same time, he discussed an anisotropic version
of the dislocation gauge theory as well as a Chern-Simons type theory
of dislocation.

\subsection{Cartan's spiral staircase as a solution of dislocation
  gauge theory}

We want to model Cartan's spiral staircase in the gauge theory of
dislocations as a homogeneous distributions of three perpendicular
forests of screw dislocations of equal strength.  Hence, for the
dislocation density, we make the ansatz
\begin{align}
\label{T-sc}
T^\a\equiv\, ^{(3)}T^\a=\AA\eta^\a\, .
\end{align}
The pitch of the helices is proportional to the constant $\AA$ with
the dimension $[\AA]=1/\text{length}$.  Thus, in the gauge
(\ref{conn0}), Eq.(\ref{Con-deco}) yields the Levi-Civita connection
\begin{equation}
\label{LC}
\widetilde{\Gamma}_{\a\b}=K_{\a\b}=-\frac{\cal A}{2}\, \eta_{\a\b}\,.
\end{equation}
and, as a consequence therefrom, the Riemannian curvature
\begin{equation}\label{Riemcurv}
\widetilde{R}^{\a\b}=-\frac{{\cal A}^2}{4}\, \vartheta^{\a\b}\,.
\end{equation}
Due to (\ref{conn0}), we find
\begin{align}
T^\a =-\widetilde{\Gamma}_\b{}^{\a}\wedge\vt^\b\, .
\end{align}
Note that (\ref{Riemcurv}) {\it deviates} from the original result
(\ref{result}) of the spiral staircase.

In order to determine the excitation, we insert (\ref{T-sc}) into the
constitutive law (\ref{const_iso}):
\begin{align}
\label{H-sc}
H_\a=a_3\AA\, \hodge\eta_\a=a_3\AA\,{\hodge\hodge}\vt_\a=
a_3\AA\,\vt_\a\, .
\end{align}
In turn, the {\it moment stress} reads
help of (\ref{momstress}):
\begin{equation}\label{momstress1}
\tau_{\a\b}=\vt_{[\a}\wedge H_{\b]}=a_3\AA\vt_{\a\b}\,.
\end{equation}
Inversion yields the components of the moment stress tensor, see
(\ref{qazq}),
\begin{equation}
\mathfrak{s}_{\a\b\g}=\mathfrak{s}_{[\a\b\g]}=a_3\AA\eta_{\a\b\g}\,.
\end{equation}
This is apparently a pure constant torque.

Turning now to the {\it force stress,} we first compute the dislocation
stress $E_\a$ which, for a quadratic Lagrangian, can be rewritten as
\begin{align}
\label{E1}
E_\a =\frac{1}{2} \left[\big(e_\a\iner T^\b\big)\wedge
H_\b-\big(e_\a\iner H_\b\big)T^\b\right]\, .
\end{align}
We substitute (\ref{T-sc}) and (\ref{H-sc}) and find
\begin{align}
\label{E2}
E_\a =\frac{1}{2}\,a_3\AA^2 \left[\big(e_\a\rfloor \eta^\b\big)\wedge
  \vt_\b-\big(e_\a\rfloor \vt_\b\big)\eta^\b\right]=\,
\frac{1}{2}\,a_3\AA^2\eta_\a\, .
\end{align}
This is a hydrostatic pressure quadratic in $\AA$, as an inversion
\`a la (\ref{qazq}) shows explicitly:
\begin{align}
\label{t-el}
\mathfrak{t}_\a{}^\b=\frac{1}{2}\, a_3 \AA^2\delta_\a^\b\, .
\end{align}
Note that $\cal A$ enters this equation quadratically.

By means of the {\it field equation} (\ref{YM-fe}) we can determine the
elastic stress $\Sigma_\a$. We first differentiate (\ref{H-sc}) and
find
\begin{equation}\label{dH}
\d H_\a=a_3\AA \,\d \vt_\a=a_3\AA^2\,\eta_\a\,.
\end{equation}
Now we turn to (\ref{YM-fe}) and substitute (\ref{dH}) and (\ref{E2})
into it:
\begin{align}
\Sigma_\a=\d H_\a-E_\a=\frac{1}{2}\, a_3\AA^2 \,\eta_\a\,.
\end{align}
Therefore, we have for both stress 2-forms
\begin{align}
\label{Sigma-el}
\Sigma_\a=E_\a=\frac{1}{2}\, a_3\AA^2 \,\eta_\a\,,
\end{align}
and the total stress adds up to
\begin{align}\label{tot}
  \Sigma_\a+E_\a= a_3\AA^2 \,\eta_\a\,.
\end{align}
If we denote the total stress tensor of the left-hand-side of
(\ref{tot}) by $^{\text{tot}}\mathfrak{t}_\a{}^\b$, compare
(\ref{qazq}), then we find again a hydrostatic pressure, namely
\begin{equation}
  ^{\text{tot}}\mathfrak{t}_\a{}^\b =-\,^{\text{tot}}p\delta_\a^\b
  =a_3\,{\cal    A}^2\,  \delta_\a^\b\,.
\end{equation}

Accordingly, collecting our results, the force and moment stresses
turn out to be
\begin{equation}\label{equi}
  ^{\text{tot}}\Sigma_\a =a_3{\cal A}^2\eta_\a\,,\qquad
  \tau_{\a\b}=a_3{\cal A}\vt_{\a\b}\,.
\end{equation}
As a check we differentiate the total force stress
\begin{equation}\label{check1}
  \d\, ^{\text{tot}}\Sigma_\a
  =a_3{\cal A}^2\d\eta_\a=-a_3{\cal A}^2\eta_{\a\b}
  \wedge T^\b=- a_3{\cal A}^3\eta_{\a\b}\wedge\eta^\b=0\,,
\end{equation}
since $\eta_{\a\b}\wedge\eta^\b=
\,\hodge\vt_{\a\b}\wedge\,\hodge\vt^\b=\vt^\b\wedge\vt_{\a\b}=0$.
Hence the force equilibrium law (\ref{forceequi}) is
guaranteed.
Moreover, even the
Peach-Koehler force 3-form~(\ref{PKF}) itself is zero:
\begin{align}
f_\alpha= (e_\a \iner T^\b)\wedge\Sigma_\b
= -\frac{1}{2}\, a_3 {\cal A}^3\eta_{\a\b}\wedge\eta^\b=0\, .
\end{align}
Thus, the elastic stress and the dislocation stress are conserved, separately,
$\d\Sigma_\a=0$ and $\d E_\a$.
Similarly, we obtain
\begin{equation}\label{check2}
  \d\tau_{\a\b}=a_3{\cal A} \d \vt_{\a\b}=a_3{\cal A}\, T_{[\a}\wedge\vt_{\b]}
=a_3 {\cal A}^2\, \eta_{[\a}\wedge\vt_{\b]}=0\,.
\end{equation}
Thus, the moment equilibrium law (\ref{momequi}) is also fulfilled since
$^{\text{tot}}\Sigma_\a$ is symmetric and $T_{[\a}\wedge H_{\b]}=0$.

If we look back to our scheme (\ref{fund1''}) to (\ref{fund6''}), then we
recognize that all equations are now fulfilled with the exception of
(\ref{fund1''}) and the constitutive law in (\ref{fund5''}). Let us first turn
to the former equation. Since the torsion is known, we can write down
(\ref{fund1''}) explicitly:
\begin{equation}\label{Tor1}
T^\a = \AA \eta^\a =\AA \,^\star \vt^\a = \d\vt^\a= (\partial_i e_j{}^\a)\, 
\d x^i\wedge \d x^j\,.
\end{equation}
Applying the star to the equation, we find
\begin{equation}\label{Tor2}
\AA \vt^\a = \left(\partial_i e_j{}^\a\right)\,^\star (\d x^i \wedge \d x^j)\,.
\end{equation}
With $^\star \left(\d x^i \wedge \d x^j\right)=\eta^{ijk} \d x_k$, we have
\begin{equation}\label{Tor3}
\AA e_i{}^\a =  \eta_{i}{}^{jk} (\partial_j e_k{}^\a)\,.
\end{equation}
In symbolic notation we can write
\begin{equation}\label{Tor4}
\AA \,\mathbf{\vt}^\a=(\text{curl}\, \mathbf{\vt})^\a \,.
\end{equation}
This means that the object of anholonomity is constant, that is, we have a
constant `vorticity' field. This coincides with our intuition of the
distribution of screw dislocations.

In order to solve (\ref{Tor3}) approximately, we linearize the coframe,
$e_i{}^\a=\delta_i^\a+h_i{}^\a+\dots$. We substitute it into (\ref{Tor3}) and
find
\begin{equation}\label{linear}
\AA \delta_i^\a =  \eta_{i}{}^{jk} (\partial_j h_k{}^\a)\,.
\end{equation}
Then we can read off the result $h_k{}^\a=\frac{\AA}{2}\eta_k{}^{\a l}x_l$
or, in terms of the distortion 1-form 
\begin{equation}
\label{linear1}
\vt^\a=\left(\delta_i^\a -\frac{\AA}{2}\eta^\a{}_{ij}x^j\right)\d x^i\,.
\end{equation}
The distortion describes a rotation perpendicular to the $(\a i)$-plane.

With the help of (\ref{linear1}) we find the three different pieces of
the distortion as
\begin{align}
\label{F-Ir}
^{(1)}\vt^\a=0\, ,\qquad 
^{(2)}\vt^\a= -\frac{\AA}{2}\eta^\a{}_{ij}x^j \, \d x^i\, ,\qquad
^{(3)}\vt^\a=\delta^\a_i \,\d x^i\, .
\end{align}
In turn, the constitutive law (\ref{const_Sigma}) expresses the stress
$\Sigma_\a$ in terms of the distortion (\ref{linear1}).
If we substitute (\ref{F-Ir}) into (\ref{const_Sigma}) and compare it with
(\ref{Sigma-el}),
we find for the Hooke type moduli
\begin{align}
\label{rel-cm}
c_2=0\, , \qquad c_3= \frac{1}{2}\, a_3 \AA^2\, .
\end{align}
The elastic modulus $c_3$ corresponds to the compression modulus
$\kappa$, with $c_3=\kappa/3$.  In this way, the dislocation modulus
$a_3$ can be expressed in terms of the modulus of compression and the
pitch of helices:
\begin{align}
a_3=\frac{2\kappa}{3\AA^2}\, .
\end{align}
Thus, we conclude that Cartan's spiral staircase is a solution in the
gauge theory of dislocations provided the moduli of the underlying
material obey the conditions (\ref{rel-cm}).

Summing up: In the framework of dislocation gauge theory, we found
\begin{align}
\label{result2}
T^\a=2\omega\, \eta^\a\, ,\qquad
R^{\a\b}=0\, ,\qquad
\widetilde{R}^{\a\b}=-\omega^2\, \vt^{\a\b}\,,\quad
{\text{with}}\quad \omega=\frac{\AA}{2}\, .
\end{align}
If we compare with Cartan's spiral staircase (\ref{result}), we
observe that there is a difference to second order in $\omega$: the
Riemann and the Riemann-Cartan curvature exchange places. For Cartan's
spiral staircase we have a teleparallelism with respect to the
Riemannian (or Levi-Civita) connection, for the forests of the screw
dislocations the underlying Riemann-Cartan space is
teleparallel. Insofar the dislocation theory led to a slightly
different result as compared to the spiral staircase. This is not
unexpected: Let us consider the case of a constant dislocations
density in a real crystal. Then dislocation theory for geometrical
reasons predicts that the underlying connection, in terms of which the
torsion is defined, has to be teleparallel, see \cite{Kroener1980}.

A second remark is in order: The torque stress $\tau_{\a\b}=a_3\AA
\vt_{\a\b}$ is linear in the pitch $\AA$ and the hydrostatic pressure
$\Sigma_\a=\frac 12 a_3\AA^2\eta_\a$ quadratic in $\AA$.  Thus, the
pressure corresponds to a nonlinear effect. This is consistent with
the screw dislocation distribution specified. In linear elasticity,
the stress fields of screw dislocations are represented by pure shear
stresses. Therefore, a constant pressure, caused by screw dilations,
can only occur in the nonlinear regime. Hence our picture is
apparently consistent.

\section{Discussion}

As we have seen, Cartan's spiral staircase corresponds to a
homogeneous and isotropic torsion distribution in {\it three}
dimensions. Is it also possible to have such torsion distributions in
{\it two} dimensions in the case of positive definite signature? No,
this is not possible since for $n=2$ from the three irreducible pieces
of the torsion only the {\it vector} (or trace) piece $^{(2)}T^\a$ survives,
see (\ref{trator}). Isotropy then requires $^{(2)}T^\a=0$.
Geometrically 2-dimensional torsion distributions have been studied by
Schuecking and Surowitz \cite{SchuckingSurowitz}, Sec.14 (see also
\cite{SchuckingTorsion,SchuckingApple,Maluf}).

Thus, we have a method to visualize a three-dimensional
distributions of homogeneous and isotropic torsion, and this may help
to understand the corresponding situations in gravitational physics,
in particular in the framework of the Poincar\'e gauge theory of
gravitation. We wonder whether one can find in this framework a simple
cosmological model\footnote{For such models one should compare, for
  instance, \cite{Nara} and \cite{Shie}.} with constant and isotropic
torsion.

\section*{Acknowledgement}
M.L.\ has been supported by an Emmy-Noether grant of the Deutsche
Forschungsgemeinschaft (Grant No. La1974/1-2, La1974/1-3). We
acknowledge helpful remarks to an earlier version of our paper by
Eckehard Mielke (Mexico City) and Engelbert Sch\"ucking (New York).

\end{document}